\documentclass{jltp}
\usepackage{graphicx}

\title{Alkali Atoms Attached to $^3$He Nanodroplets}
\author{
R. Mayol$^*$, F. Ancilotto$^\dag$, M. Barranco$^*$, O. B\"unermann$^\ddag$, \\
M. Pi$^*$, and F. Stienkemeier$^\ddag$}
\address{$^*$Departament ECM, Facultat de F\'{\i}sica.\\
Universitat de Barcelona, E-08028 Barcelona, Spain.\\
$^\dag$INFM (Udr Padova and DEMOCRITOS National Simulation
Center, Trieste); \\ Dipartimento di Fisica ``G. Galilei'',
Universit\`a di Padova I-35131 Padova, Italy\\
$^\ddag$Fakult\"at f\"ur Physik, Universit\"at Bielefeld, D-33615
Bielefeld, Germany\\}

\runninghead
{R. Mayol {\it et al.}}
{Alkali Atoms Attached to $^3$He Nanodroplets}

\begin{document}

\maketitle

\begin{abstract}

We have experimentally studied the electronic  $3p\leftarrow 3s$
excitation of Na atoms attached to $^3$He droplets by means of
laser-induced fluorescence as well as beam depletion spectroscopy. From
the similarities of the spectra (width/shift of absorption lines) with
these of Na on $^4$He droplets, we conclude that sodium atoms reside in
a ``dimple'' on the droplet surface and that superfluid-related effects
are negligible. The experimental results are supported by Density
Functional calculations at zero temperature, which confirm the surface
location of Na, K and Rb atoms on $^3$He droplets. In the case of Na,
the calculated shift of the excitation spectra for the two isotopes is
in good agreement with the experimental data.

\vspace*{0.5cm}
\noindent PACS 68.10.-m, 68.45.-v, 68.45.Gd

\end{abstract}

\section{INTRODUCTION}

Detection of laser-induced fluorescence (LIF) and beam depletion (BD)
signals upon laser excitation provides a sensitive spectroscopic
technique to investigate electronic transitions of chromophores attached
%to $^4$He nanodroplets.\cite{stienke2} While most of atomic and
to $^4$He droplets.\cite{stienke2} While most of atomic and
molecular dopants
% frank
%
% migrate to the center of the droplet,
%
%% some theoreticians don't like that because as in the case of an
%% harmonic oscillator the center position is not very likely
%
submerge in helium,
% frank
alkali atoms (and alkaline earth atoms to some extent\cite{Sti:1997b})
have been found to reside on the surface of $^4$He droplets, as
evidenced by the much narrower and less shifted spectra when compared
to those found in bulk liquid
$^4$He.\cite{scoles,ernst1,stienke3,Ernst:2001a} This result has been
confirmed by Density Functional (DF)\cite{anci1} and Path Integral
Monte Carlo (PIMC)\cite{nakayama} calculations, which predict surface
binding energies of a few Kelvin, in agreement with the measurements of
detachment energy thresholds using the free atomic emissions.\cite{KKL}
The surface of liquid $^4$He is only slightly perturbed by the presence
of the impurity, which produces a ``dimple'' on the underlying liquid.
The study of these states can thus provide useful information on
surface properties of He nanodroplets complementary to that supplied by
molecular-beam scattering experiments.\cite{Dal98,har01}
% frank
Hence, alkalis on the surface of helium droplets are ideal probes
%
%The behavior of dopants in He clusters is especially appealing. In
%particular, probes at the surface of the droplets are desirable because
%they allow
% frank
to investigate the liquid--vacuum interface as well as droplet surface
excitations.
% frank

% frank
Microscopic calculations of $^3$He droplets are
scarce.\cite{panda,gua00} The properties of $^3$He droplets doped with
some inert atoms and molecular impurities have been addressed within
the Finite Range Density Functional (FRDF) theory,\cite{gar98} that has
proven to be a valuable alternative to Monte Carlo methods which are
notoriously difficult to apply to Fermi systems. Indeed, a quite
accurate description of the properties of inhomogeneous liquid $^4$He
at zero temperature ($T$) has been obtained within  DF
theory,\cite{prica} and a similar approach has followed for $^3$He (see
Ref. \onlinecite{gar98} and Refs.~therein).

\section{RESULTS}

\begin{figure}
% frank
%\centerline{\includegraphics[width=8cm,clip]{fig1.eps}}
\centerline{\includegraphics[width=10cm,clip]{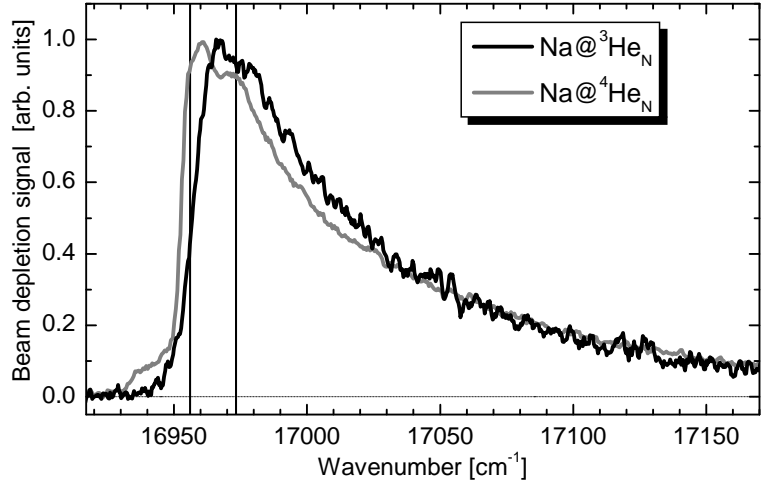}}
% frank
\caption[] {Beam depletion spectra of Na atoms attached to
$^3$He/$^4$He nanodroplets. The vertical lines indicate the positions
of the two
% frank
fine structure
% frank
components of the Na gas-phase  $3p\leftarrow 3s$ transition.}
\label{exp_spectra}
\end{figure}

The experiments we report have been performed in a helium droplet
machine used earlier for LIF and BD studies, and is described
elsewhere.\cite{Sti:1997b} Briefly, helium gas is expanded under
supersonic conditions from a cold nozzle forming a beam of droplets
traveling freely under high vacuum conditions. The droplets are doped
downstream employing the pick-up technique: in a heated scattering
cell, bulk sodium is evaporated in such a way that, on average, a
single metal atom is carried by each droplet. Since electronic
excitation of alkali-doped helium droplets is eventually followed by
desorption of the chromophore, BD spectra can be registered by a
Langmuir-Taylor surface ionization detector.\cite{Sti:2000b}
Phase-sensitive detection with respect to the chopped laser or droplet
beam was used. For that reason the BD signal
(cf.~Fig.~\ref{exp_spectra}), i.e.~a decrease in intensity, is directly
recorded as a positive yield. For these experiments, a new droplet
source was built to provide the necessary lower nozzle temperatures to
condense $^3$He droplets.

For the spectroscopic measurements presented in the following, we have
set the source pressure to 20\,bar and the nozzle temperature to 11\,K
for  $^3$He, and to 15\,K for $^4$He. These conditions are expected to
result in comparable mean cluster sizes around 5000 atoms per
droplet.\cite{Toe:unpublished,har01} In Fig.~\ref{exp_spectra} the
absorption  spectrum of Na atoms attached to $^3$He nanodroplets is
shown in comparison to Na-doped $^4$He droplets.

The outcome of the spectrum of Na attached to $^3$He nanodroplets is
very similar to the spectrum on $^4$He droplets. The asymmetrically
broadened line is almost unshifted with respect to the gas-phase
absorption. This absence of a shift immediately confirms the surface
location because atoms embedded in bulk superfluid helium are known to
evolve large blue-shifts of the order of a couple of hundreds of
wavenumbers and much more broadened absorption
lines.\cite{Takahashi:1993} A blue shift is a consequence of the
repulsion of the helium environment against the spatially enlarged
electronic distribution of the excited state (``bubble effect''). The
interaction towards the $^3$He droplets appears to be slightly
enhanced, evidenced by the small extra blue shift of the spectrum
compared to the $^4$He spectrum. In a simple picture this means that
more helium atoms are contributing or, in other words, a more prominent
``dimple'' interacts with the chromophore. The upper halves of the
spectra are almost identical, when shifting the $^3$He spectrum by
$7.5\pm 1$\,cm$^{-1}$ to lower frequencies.

\begin{figure}[tbh!] % Figure 2
\centerline{\includegraphics[width=8cm,angle=270,clip]{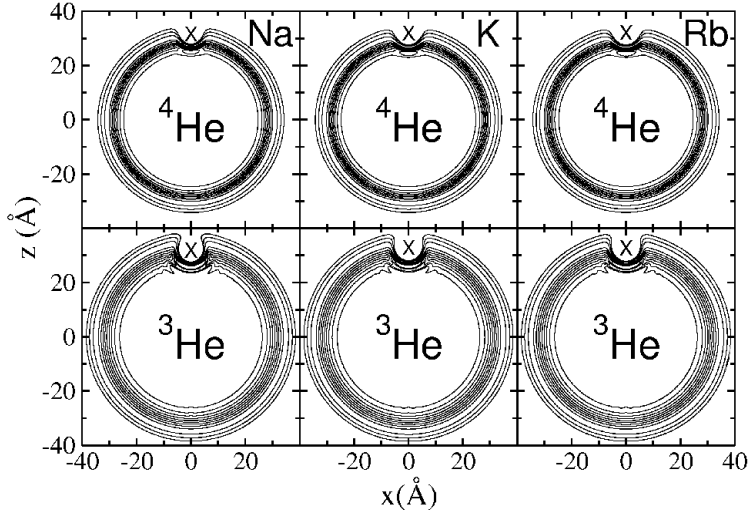}}
\caption[]
{Equidensity lines in the $x-z$ plane showing the stable state
of an alkali atom (cross) on a He$_{2000}$ droplet. The 9 inner lines
correspond to densities $0.9 \rho_0$ to $0.1 \rho_0$, and the 3 outer
lines to $10^{-2} \rho_0$, $10^{-3} \rho_0$, and  $10^{-4} \rho_0$
($\rho_0=0.0163$ \AA$^{-3}$ for $^3$He, and 0.0218 \AA$^{-3}$ for
$^4$He). }
\label{fig2}
\end{figure}

FRDF calculations at $T=0$ confirm the picture emerging from the
measurements, i.e.~the surface location of Na on $^3$He nanodroplets
causing a more pronounced ``dimple''  than in $^4$He droplets. We have
investigated the stable configurations of an alkali atom on both $^3$He
and $^4$He clusters of different sizes. The FRDF's used for $^3$He and
$^4$He are described in Refs. \onlinecite{bar97} and
\onlinecite{may01}. The large number of $^3$He atoms we are considering
allows to use the extended Thomas-Fermi approximation.\cite{TF} The
presence of the foreign impurity is modeled by a suitable potential
obtained by folding the helium density with an alkali-He pair
potential. We have used the potentials proposed by Patil\cite{patil} to
describe the impurity-He interactions. Fig.~\ref{fig2} shows the
equilibrium configuration for alkali atoms adsorbed onto He$_{2000}$
clusters. Comparison with the stable state on the $^4$He$_{2000}$
cluster shows that, in agreement with the experimental findings
presented before, the ``dimple'' structure is more pronounced in the
case of $^3$He, and that the alkali impurity lies {\it inside} the
surface region for $^3$He and {\it outside} the surface region for
$^4$He (the surface region is usually defined as that comprised between
the radii at which $\rho=0.1 \rho_0$ and $\rho=0.9 \rho_0$, where
$\rho_0$ is the He saturation density\cite{Dal98,har01,TF}). This is
due to the lower surface tension of $^3$He as compared to that of
$^4$He, which also makes the surface thickness of bulk liquid and
droplets larger for $^3$He than for $^4$He.\cite{Dal98,har01}

The deformation of the surface upon alkali adsorption is
characterized
% frank
%\cite{anci1}
% frank
by the ``dimple'' depth, $\xi$, defined as the difference between the
position of the dividing surface at $\rho \sim \rho_0/2$, with and
without impurity,
% frank
respectively.\cite{anci1}
% frank
For $^3$He we have found $\xi \sim$4.4, $\sim$4.1, and $\sim$4.3 \AA,
for Na, K and Rb, respectively. The corresponding values for $^4$He are
$\xi \sim$2.3, $\sim$2.3, and $\sim$2.0 \AA, respectively.

\begin{figure}[tbh!] % Figure 3
\centerline{\includegraphics[width=8cm,angle=270,clip]{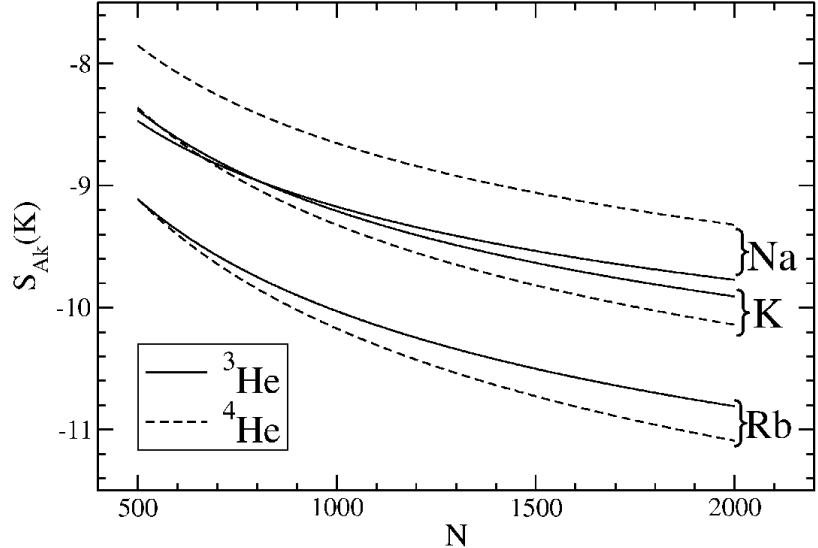}}
\caption[]
{ Solvation energies as a function of the number of atoms in
the droplet.
 }
\label{fig3}
\end{figure}

Solvation energies  defined as $S_{Ak}=E({\rm Ak}@{\rm He}_N)-E({\rm
He}_N)$ have also been calculated and are shown in Fig. \ref{fig3}.
%. For $^3$He$_{2000}$ we have found
%-9.77, -9.91, and -10.81 K
% frank
%
% I would suggest to drop the last digit or does one trust numbers to
% 1/100 K ? It is up to you.
%
% frank
%for Na, K and Rb, respectively. The corresponding values for
%$^4$He$_{2000}$ are -9.32, -10.14, and -11.09 K, respectively.
The
value $S_{Na} \sim -12$ K has been obtained within FRDF theory for Na
adsorbed on the {\it planar} surface of $^4$He (Ref.
\onlinecite{anci1}), which corresponds to the $N=\infty$ limit.
%
%    Manuel
%    Please, pay attentin to the next sentence!!!!
%
A detailed discussion on solvation energies and
AkHe-exciplex formation on helium nanodroplets will be presented
elsewhere.

Finally, we have obtained the shift between the $^3$He and $^4$He
spectra in Fig.~\ref{exp_spectra} within the Frank-Condon
approximation, i.e. assuming that the ``dimple'' shape does not change
during the Na excitation. The shift is calculated within the model
given in Ref. \onlinecite{Kan94}, evaluating Eq.~6 therein, both for
$^3$He and $^4$He. We have used the excited state A $^2\Pi$ and B
$^2\Sigma$ potentials of Ref. \onlinecite{nakayama} because their Na-He
% frank
% GS
ground state
% frank
potential is very similar to the Patil potential we have used to obtain
the equilibrium configurations. For $N=2000$,  we find that the $^3$He
spectrum is blue-shifted with respect to the $^4$He one by
6.4\,cm$^{-1}$, in good agreement with the experimental value of
$7.5\pm 1$\,cm$^{-1}$ as extracted from Fig.~\ref{exp_spectra}.

Our results thus show that alkali adsorption on $^3$He droplets occurs
in very much the same way as in the case of $^4$He, i.e., the adatom is
located on the surface, though in a slightly more pronounced
``dimple''. The similarities in the experimental spectra are certainly
remarkable for two apparently very different fluids, one normal and the
other superfluid, and clearly indicate that superfluidity does not play
any substantial role in the processes described here (we recall that
while $^4$He droplets, which are detected at an experimental $T$ of
$\sim$ 0.38\, K, are superfluid, these containing only $^3$He atoms,
even though detected at a lower $T$ of $\sim$ 0.15\,K, do not exhibit
superfluidity\cite{grebenev}). This is likely a consequence of the very
fast time scale characterizing the Na electronic excitation compared to
that required by the He fluid to readjust. The excitation occurs in a
``frozen'' environment and the only significant difference between
$^3$He and $^4$He is due to the different structure of the``dimple'',
which accounts for the small shift in their spectra observed in the
experiments and found in our calculations as well.

\section*{ACKNOWLEDGMENTS}

We thank Flavio Toigo for useful comments. This work has been supported
by grants MIUR-COFIN 2001 (Italy), BFM2002-01868 from DGI (Spain), and
2001SGR-00064 from Generalitat of Catalunya as well as the DFG
(Germany).


\begin{thebibliography}{99}

\bibitem{stienke2}F. Stienkemeier and A.F. Vilesov,
J. Chem. Phys. {\bf 115}, 10119 (2001).

\bibitem{Sti:1997b}
F. Stienkemeier, F. Meier, and H.~O. Lutz, J. Chem. Phys. {\bf 107},
10816 (1997); Eur. Phys. J. D {\bf 9}, 313 (1999).

\bibitem{scoles} F. Stienkemeier et al., Z. Phys. D
{\bf 38}, 253 (1996).

\bibitem{ernst1} F. Stienkemeier et al.,
J. Chem. Phys. {\bf 102}, 615 (1995);

\bibitem{stienke3}
C. Callegari et al., J. Phys. Chem. {\bf 102}, 95 (1998).

\bibitem{Ernst:2001a}
F. Br\"uhl, R. Trasca, and W. Ernst, J. Chem. Phys. {\bf 115},  10220
(2001).

\bibitem{anci1} F. Ancilotto et al., Z. Phys. B
{\bf 98}, 323 (1995).

\bibitem{nakayama} A. Nakayama and K. Yamashita,
J. Chem. Phys. {\bf 114}, 780 (2001).

\bibitem{KKL} J. Reho et al., Faraday Discuss \textbf{108}, 161 (1997).

\bibitem{Dal98}
F. Dalfovo, J. Harms, and J.P. Toennies, Phys. Rev. B
{\bf 58}, 3341 (1998).

\bibitem{har01} J. Harms et al., Phys. Rev. B {\bf 63}, 184513 (2001).

\bibitem{panda}V.R. Pandharipande, S.C. Pieper, and
R.B. Wiringa, Phys. Rev. B {\bf 34}, 4571 (1986).

\bibitem{gua00}
R. Guardiola, Phys. Rev. B {\bf 62}, 3416 (2000).

\bibitem{gar98} F. Garcias et al., J. Chem. Phys. {\bf 108},
9102 (1998); {\it ibid.} {\bf 115}, 10154 (2001).

\bibitem{prica} F. Dalfovo et al., Phys. Rev. B
{\bf 52}, 1193 (1995).

\bibitem{Sti:2000b}
F. Stienkemeier et al.,  Rev. Sci. Instr. {\bf 71}, 3480  (2000).

\bibitem{Toe:unpublished}
J. Harms and J.~P. Toennies, unpublished results.

%\bibitem{Lehmann:2000c}
%J. Reho et al., J. Chem. Phys. {\bf 113},  9686  (2000).

\bibitem{Takahashi:1993}
Y. Takahashi et al., Phys. Rev. Lett. {\bf 71}, 1035  (1993).

\bibitem{bar97} M. Barranco et al.,
Phys. Rev. B {\bf 56}, 8997 (1997).

\bibitem{may01} R. Mayol et al., Phys. Rev. Lett.
{\bf 87}, 145301 (2001).

\bibitem{TF}
S. Stringari and J. Treiner, J. Chem. Phys. {\bf 87},
5021 (1987).

\bibitem{patil} S.H. Patil, J. Chem. Phys. {\bf 94}, 8089 (1991).

\bibitem{Kan94}
S.I. Kanorsky  al., Phys. Rev. B {\bf 50}, 6296 (1994).

\bibitem{grebenev} S. Grebenev, J.P. Toennies, and A.F. Vilesov,
Science {\bf 279}, 2083 (1998).

\end{thebibliography}
\end{document}